\documentclass[aps,prl,reprint,superscriptaddress,floatfix,amsmath,showpacs]{revtex4-1}

\usepackage{graphicx}
\usepackage{amsmath}
\usepackage{textcomp} 

\begin{document}

\title{Spin pumping by parametrically excited exchange magnons}

\author{C.~W.~Sandweg}
\email{sandweg@physik.uni-kl.de}
\affiliation{Fachbereich Physik and Forschungszentrum OPTIMAS, Technische Universit\"at Kaiserslautern, 67663
Kaiserslautern, Germany}

\author{Y.~Kajiwara}
\affiliation{Institute for Materials Research, Tohoku University, Sendai 980-8577, Japan}

\author{A.~V.~Chumak}
\affiliation{Fachbereich Physik and Forschungszentrum OPTIMAS, Technische Universit\"at Kaiserslautern, 67663
Kaiserslautern, Germany}

\author{A.~A.~Serga}
\affiliation{Fachbereich Physik and Forschungszentrum OPTIMAS, Technische Universit\"at Kaiserslautern, 67663
Kaiserslautern, Germany}

\author{V.~I.~Vasyuchka}
\affiliation{Fachbereich Physik and Forschungszentrum OPTIMAS, Technische Universit\"at Kaiserslautern, 67663
Kaiserslautern, Germany}

\author{M.~B.~Jungfleisch}
\affiliation{Fachbereich Physik and Forschungszentrum OPTIMAS, Technische Universit\"at Kaiserslautern, 67663
Kaiserslautern, Germany}

\author{E.~Saitoh}
\affiliation{Institute for Materials Research, Tohoku University, Sendai 980-8577, Japan}
 \affiliation{CREST, Japan Science and Technology Agency, Sanbancho, Tokyo 102-0075, Japan}
 \affiliation{Advanced Science Research Center, Japan Atomic Energy Agency, Tokai 319-1195, Japan}

\author{B.~Hillebrands}
\affiliation{Fachbereich Physik and Forschungszentrum OPTIMAS, Technische Universit\"at
Kaiserslautern, 67663 Kaiserslautern, Germany}

\date{\today}

\begin{abstract}

We experimentally show that exchange magnons can be detected using a combination of spin pumping and inverse spin-Hall effect (ISHE) proving its wavelength integrating capability down to the sub-micrometer scale. The magnons were injected in a ferrimagnetic yttrium iron garnet film by parametric pumping and the ISHE-induced voltage was detected in an attached Pt layer. The role of the density, wavelength, and spatial localization of the magnons for the spin pumping efficiency is revealed.

\end{abstract}

\maketitle

Spintronics, the field of spin-based data storage and processing, is a very promising candidate to overcome the limits of conventional charge-based electronics \cite{spintronic}. The spin pumping effect turned out to be an important mechanism for the generation of a spin current in nonmagnetic conductors \cite{Tserkovnyak, Mizukami}. The inverse spin-Hall effect (ISHE) can subsequently be used to convert this spin current into a detectable charge current \cite{Saitoh-2006}. It has been shown recently that it is even possible to develop magnetic insulator based spintronic devices in which the information is carried by magnons, the quanta of spin waves, instead of spin-polarized electrons \cite{Kajiwara, Uchida2010}.

Two different ways are often used to excite magnons. The first one is force excitation using a microwave magnetic field with the same frequency as the spin wave. With this method the wavelength of the excited spin waves is restricted by the lateral size of the antenna and mostly dipolar magnons with small wavenumbers (i.e. long wavelengths) and frequencies near the ferromagnetic resonance (FMR) can be excited \cite{GangulyWebb,YIG_magnonics}. The second one is parametric pumping where the spin waves are excited at half of the frequency of the microwave magnetic field \cite{YIG_magnonics, Melkov, Lvov}. In the case of parametric pumping, there is no upper limit for the wavenumbers of the excited spin waves. This allows the excitation of exchange magnons with short wavelengths.
To our best knowledge, in all previous experiments (e.g. \cite{Mizukami, Saitoh-2006, Kajiwara, Sandweg-2010, Ando-2009, Costache}) the spin pumping mechanism and the ISHE has been investigated only for the uniform precession or dipolar spin waves with long wavelengths and frequencies near the FMR.

\begin{figure}[ht]
\includegraphics {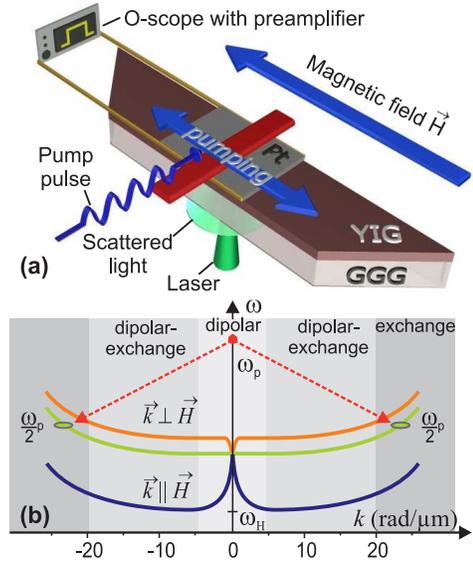}
\caption{\label{Fig1} (Color online) (a) Sketch of the experimental setup.
(b) Schematic illustration of the magnon spectrum and the mechanism of the parametric pumping process: one quantum of the pump field of $\omega_\mathrm{p}$ frequency splits into two magnons of $\omega_\mathrm{p}/2$ frequency.}
\end{figure}

In this letter we present the spin pumping effect by exchange magnons and its detection via the ISHE voltage in a ferrite/platinum bi-layer structure. The magnons were excited in a ferrite film using the parametric mechanism. Brillouin light scattering (BLS) spectroscopy was used to identify the  excited magnons. By varying the value of the magnetizing field we were able to change the wavelength of the magnons down to sub-micron values. The exchange magnons are found to contribute significantly to the spin pumping process.

Figure~\ref{Fig1}(a) shows a sketch of the experimental setup. The sample used in the present study comprises a 2.1~$\mu$m thick single-crystal ferrite yttrium iron garnet (YIG, Y$_\mathrm{3}$Fe$_\mathrm{5}$O$_\mathrm{12}$) film grown on a gallium gadolinium garnet (GGG, Gd$_\mathrm{3}$Ga$_\mathrm{5}$O$_\mathrm{12}$) substrate by liquid phase epitaxy. A 10~nm thick $3 \times 3$~mm$^{2}$ platinum (Pt) layer is deposited directly onto the YIG film. The Pt pad is wired to a voltage preamplifier and an oscilloscope in order to detect the electromotive force $V_\mathrm{ISHE}$ generated by the ISHE. The magnetizing field $\vec{H}$ is applied in the standard orientation to observe the ISHE \cite{Saitoh-2006, Kajiwara, Sandweg-2010} so that the electron carried spin current (which propagates into the Pt layer from the YIG/Pt interface) and the ISHE induced charge current are perpendicular to each other and to the field $\vec{H}$. Magnons contributing to the ISHE voltage are parametrically pumped by a microwave Oersted field induced with a 50~$\mu$m-wide microstrip, which is insulated from the Pt layer by a thin non-magnetic dielectric coating of cyan-acrylate. Both parallel and perpendicular (relative to $\vec{H}$) components of the Oersted field contribute to the parametric pumping as it has been shown in Ref.~\cite{Timo}. The pump frequency was held constant at 14~GHz and the pump power $P$ was varied between 1.7~W and 28.2~W. In order to reduce the microwave heat we used 5~$\mu$s-long microwave pulses with 50~$\mu$s repetition time rather than continuous waves. The magnons were detected using BLS spectroscopy: the probing light beam was focused on the YIG/Pt sample and the inelastically scattered light, whose intensity is proportional to the quantity of magnons, was analyzed with a tandem Fabry-P\'{e}rot interferometer \cite{Demokritov2001}.

Figure~\ref{Fig1}(b) schematically illustrates the magnon spectrum (frequency $\omega$ vs wavenumber $k$) in a YIG film and the mechanism of parametric electromagnetic pumping. The entire spectrum, which comprises all possible directions of magnon propagation, can be separated in three different regions. In the dipolar area, where magnons have small wavenumbers the dipolar interaction is dominant. In a region of $k$ from $1$~rad/$\mu$m to $10$~rad/$\mu$m the magnon dispersion relations are influenced by both dipolar and exchange interactions. In the third area (above $10$~rad/$\mu$m), only exchange magnons exist.

\begin{figure}[ht]
\includegraphics {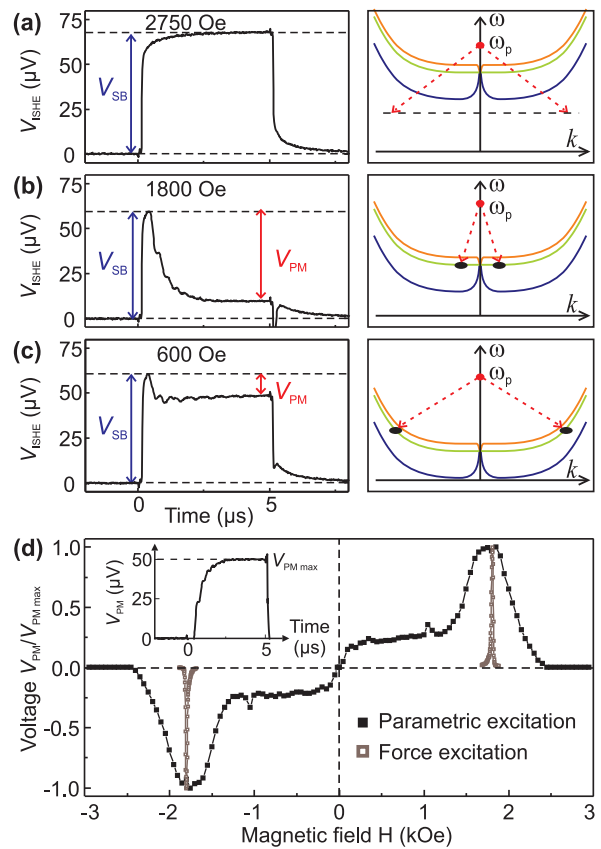}
\caption{\label{Fig2} (Color online) (a)-(c) Waveforms of the electromotive force signal $V_\mathrm{ISHE}$ (left column) and pumping schemes for different magnetizing fields $H$. (d) Normalized spin pumping induced voltage dependencies $V_\mathrm{PM} (H)$ for parametrically injected magnons (black filled squares) and force excited magnons (gray empty squares). The inset shows the time profile of the pure $V_\mathrm{PM}$ signal without spin-Seebeck effect contribution $V_\mathrm{SB}$ at $H=1800$~Oe.}
\end{figure}

The electromagnetic pumping used in our experiment can be described in terms of energy quanta, where a single microwave photon (associating with the parallel component of the microwave Oersted field) or magnon (nonresonantly excited  by the perpendicular component of the microwave field) with the frequency $\omega_\mathrm{p}$ and near zero wavenumber splits into two magnons with $\omega_\mathrm{p}/2$ but opposite wavevectors $\vec{k}$ and $-\vec{k}$.
Due to the degeneracy in $\omega, k$-space, different magnon groups with the same frequency $\omega_\mathrm{p}/2$ are pumped at the same time, but only one survives \cite{Melkov,Lvov}. This group, the dominant group, is characterized by the lowest damping and the highest coupling to the pump field \cite{Melkov,Lvov,Timo}. The dispersion curve corresponding to the dominant group is shown by the middle green line in Fig.~\ref{Fig1}(b). The variation of the magnetizing field $H$ at constant pump frequency results in a shifting of the whole spectrum up or down and in the tuning of the excited magnons wavenumbers (see right panels in Fig.~\ref{Fig2}). In the presented experiment, the field $H$ was varied in the range from $-3000$~Oe to 3000~Oe allowing the excitation of both dipolar and exchange magnons.

The signal of the electromotive force $V_\mathrm{ISHE}$ measured for $P=8.9$~W at $H=2750$~Oe is presented in Fig.~\ref{Fig2}(a). On the right hand-side, the sketch of the corresponding pumping process demonstrates that $\omega_\mathrm{p}$/2 lies below the spin-wave spectrum and no parametric excitation can take place. In spite of this fact, a $V_\mathrm{ISHE}$-signal is detectable for this magnetic field. This voltage $V_\mathrm{SB}$ is associated with the longitudinal spin-Seebeck (SSE) effect which has recently been demonstrated for magnetic insulators \cite{Uchida2010_2}. It is independent of the value of the bias magnetic field $H$ and changes its polarity at $H=0$. A temperature gradient $\nabla T$ perpendicular to the YIG surface (and parallel to the spin current) is created due to the heating of the Pt layer by eddy currents induced by the microwave pump field. The longitudinal SSE in insulator/metal systems can be explained with an imbalance between an effective magnon temperature in YIG and an effective conduction-electron temperature in the attached Pt layer \cite{Uchida2010_2,Xiao}. The significant higher conduction electron temperature in the Pt layer leads to thermal fluctuations and an ejection of spins to the YIG.

At $H=1800$~Oe, as shown in Fig.~\ref{Fig2}(b), the situation is different since the $\omega_\mathrm{p}/2$-frequency lies above the bottom of the magnon spectrum $\omega_\mathrm{H}$ (see Fig.~\ref{Fig1}(b)), and excitation of the dipolar magnons is allowed. The time-profile of the electromotive force $V_\mathrm{ISHE}$ shows two opposing contributions which compensate each other partially. The first contribution $V_\mathrm{SB}$ belongs to the longitudinal spin-Seebeck effect, as in the former case. However, the second contribution $V_\mathrm{PM}$ can be attributed to the spin pumping by parametrically excited magnons. In this process, spins are injected into the Pt layer while in the longitudinal SSE spins are ejected from the Pt layer to the YIG. Thus, these two effects show different polarities in the signal of the electromotive force $V_\mathrm{ISHE}$. In addition, the voltage signal $V_\mathrm{PM}$ of the parametrically excited magnons is temporarily retarded since the equilibrium has to be established in the magnon system after the pump pulse is switched on. By subtracting the two contributions to the $V_\mathrm{ISHE}$-signal, the pure $V_\mathrm{PM}$ signal can be extracted (see inset Fig.~\ref{Fig2}(d)).

Figure~\ref{Fig2}(c) shows the situation for $H=600$~Oe, where only magnons in the exchange area are injected. The corresponding $V_\mathrm{ISHE}$-profile shows again the two contributions to the electromotive force. The voltage $V_\mathrm{PM}$ due to the spin pumping by exchange magnons is clearly observable.

The extracted normalized voltage $V_\mathrm{PM}$ is presented in Fig.~\ref{Fig2}(d) for every measured magnetic field $H$ from $-3000$~Oe to 3000~Oe. As expected the curve is anti-symmetrical with respect to the polarity of the magnetic field $H$. For small values of $H$ the voltage goes to zero due to the sample demagnetization and formation of domains. For large magnetizing fields ($|H|>2400$~Oe) the voltage is zero as $\omega_\mathrm{p}/2$ lies below the magnon spectrum and no parametric excitation can take place. The pronounced voltage maxima are visible around $H_\mathrm{c}=\pm 1780$~Oe, where the minima of parametric generation threshold are observed and the magnon injection is most effective.

The field $H_\mathrm{c}$ usually corresponds to the situation when the $\omega_\mathrm{p}/2$ frequency is in the vicinity of the FMR frequency of the YIG film. In order to prove this, an additional experiment has been performed: long-wavelength dipolar spin waves were excited using the same microstrip by the microwave signal of $\omega_\mathrm{s} = \omega_\mathrm{p}/2=7$~GHz frequency and ISHE voltage was measured (see the narrow peaks in Fig.~\ref{Fig2}(d)). The maxima of both dependencies coincide well. In addition, this measurement clearly demonstrates that the parametric technique allows magnon injection in a much wider wavelength range than the direct force excitation.

\begin{figure}[ht]
\includegraphics {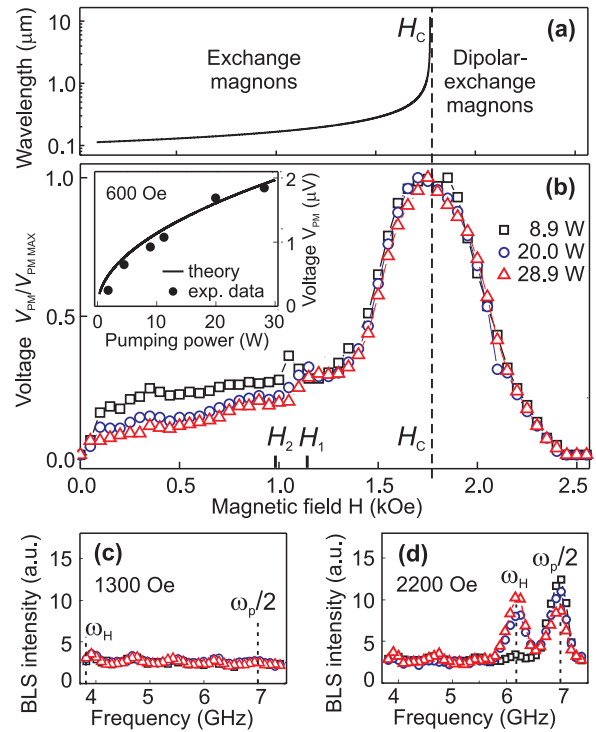}
\caption{\label{Fig3} (Color online) (a) Calculated wavelength of the parametrically injected magnons. (b) Normalized dependencies $V_\mathrm{PM} (H)$ for different pump powers. The corresponding BLS spectra for two magnetic fields are shown in panels (c) and (d). The inset in (b) shows the ISHE-signal of the parametrically excited magnons at 600~Oe for different pumping powers in comparison to the calculated power dependence.}
\end{figure}

The relatively small voltages $V_\mathrm{PM}$ in the exchange region ($|H|<1300$~Oe) in comparison to the dipolar region ($1300<|H|<2400$~Oe) can be understood by analyzing the spatial localization of the dominant group taking into account the interface nature of the spin pumping effect \cite{Tserkovnyak, Kajiwara, Sandweg-2010}. This group is characterized by the smallest magnon damping, which is mostly determined by two-magnon scattering on inhomogeneities and impurities of the sample \cite{Melkov, Sparks}. In single-crystal YIG films, these impurities are mainly localized close to the film surfaces. Thus, we may conclude that the dominant group is located in the middle of the YIG film where the scattering is the smallest. The localization of these magnons depends on their wavelengths and it determines the intensity of the magnetization precession at the YIG/Pt interface. The shorter the wavelength, the higher the localization, and consequently the spin pumping is smaller due to the decreasing of the magnetization precession on the interface.

The calculated wavelength dependence of the excited magnons as a function of the applied bias magnetic field (see Eq.~7.9 in \cite{Melkov}) is shown in Fig.~\ref{Fig3}(a). One can see that the curve follows the voltage dependence shown in Fig.~\ref{Fig2}(d) and Fig.~\ref{Fig3}(b) qualitatively: in the exchange region the wavelength as well as the ISHE-induced voltage do not change much. However,  the increase of the field in the region $\mathrm{1300~Oe} < H < H_\mathrm{c}$ results in a fast increase of the magnon wavelength as well as the detected voltage. We interpret this behavior as a direct proof that the magnon wavelength and therefore the magnon localization are of crucial importance for the spin pumping efficiency.

Note, that a slight regular increase of the parametric generation threshold ($\simeq 4$~dB from $H_\mathrm{c}$ to 250~Oe \cite{Timo}) can also partially contribute to the observed voltage decrease in the exchange region (Fig.~\ref{Fig3}(b)). However, for $H>H_\mathrm{c}$ the threshold increases rapidly up to infinity at $\omega_\mathrm{p}/2 = \omega_\mathrm{H}$ \cite{Melkov, Timo}. This effect defines the voltage fall in the high field region.

The normalized field dependence of the ISHE-induced voltage for three different pump powers are shown in Fig.~\ref{Fig3}(b). One can see that the increase of the pump power and consequently the density of the parametrically injected magnons does not change qualitatively the voltage behavior. The small increase of the voltage slope in the exchange region can be associated with a nonlinear damping caused by three-magnon scattering processes which for the given experimental conditions are allowed for fields smaller than the critical field $H_1=1150$~Oe \cite{Lvov, Falkovich}.

It is interesting that small peaks in the ISHE voltage, which are clearly visible in Fig.~\ref{Fig2}(d), appear just below the field $H_1$. Thus, they can be associated with the confluence of two parametrically injected magnons. Such a process occurs in the vicinity of the second critical field $H_2$ \cite{Lvov, Falkovich} (970~Oe for our experimental conditions). The angular momentum of one of the confluencing magnons is not conserved and must be passed to the entire sample lattice. One can assume that this angular momentum may be directly passed to a free electron in the Pt layer. As a result the spin polarization of the electron gas will increase leading to the increase of the ISHE voltage.

The dependence of the ISHE-voltage as a function of the pump power for the exchange region ($H = 600$~Oe) is shown in the inset in Fig.~\ref{Fig3}(b). In order to understand this behavior the nonlinear four-magnon scattering theory has to be taking into account \cite{Lvov}. The magnon density of the dominant group in the saturation regime is given by: $n = \sqrt{(h_\mathrm{p} V_k)^2 - \Gamma_k^2}/S$, where $S$ is a coefficient of the four-magnon scattering,  $V_k$ describes the efficiency of parametric coupling of a magnon group and the pump field, and $\Gamma_k$ is the magnon damping. As $V_\mathrm{PM}$ is proportional to the magnon density, the experimental pump power $P_\mathrm{p}$ is proportional to $(h_\mathrm{p} V_k)^2$, and the threshold of parametric generation $P_\mathrm{p~thr}$ (which is known from the experiment to be $25$~mW) is proportional to $\Gamma_k^2$ we may rewrite this equation as $V_\mathrm{PM} \propto \sqrt{P_\mathrm{p} - P_\mathrm{p~thr}}$. The theoretical dependence $V_\mathrm{PM}(P_\mathrm{p})$ shown in the inset in Fig.~\ref{Fig3}(b) is in excellent agreement with the experiment.

The parametric pumping process results not only in the increase of the density of the dominant magnon group. Due to the four-magnon scattering the redistribution of magnons over the whole magnon spectrum occurs. When the magnon density is sufficiently high, the condensation of magnons to the lowest energy state $\omega_\mathrm{H}$ in the dipolar-exchange spectral area can occur \cite{BEC}. In order to check that this process does not influence our results an additional BLS characterization of the magnon gas was performed for different pump powers. The obtained data are presented in Fig.~\ref{Fig3} for two different values of the magnetic field. In the case when the magnons are injected near the bottom of the magnon spectrum ($H=2200$~Oe) one sees the pump power dependent redistribution of the injected magnons to the lowest energy states (Fig.~\ref{Fig3}(d)). As it is visible from Fig.~\ref{Fig3}(b) this redistribution practically does not influence the ISHE voltage.
This is both due to the conservation of the total number of magnons in the four-magnon scattering process and the conservation of the magnon localization (the process develops in the dipolar-exchange region exclusively). For the small magnetizing field $H = 1300$~Oe when the exchange magnons are injected, no magnon redistribution is detected (see the absence of any signal at $\omega_\mathrm{H}$ in Fig.~\ref{Fig3}(c)). (Note, that the signal at $\omega_\mathrm{p}/2$ is not observable as the wavenumber of the injected magnons $30$~rad/$\mu$m overcome the wavenumber limitation of the BLS setup). This result shows that the only contribution to the spin pumping process comes from exchange magnons.

Concluding, we demonstrate that exchange magnons of sub-micron wavelengths with no associated dipolar field significantly contribute to the spin pumping in magnetic/nonmagnetic bi-layers. It is found that the spin pumping efficiency is mainly defined by localization of the injected magnons relative to the interlayer interface. The results are useful for the understanding of the physics of the spin pumping phenomenon and are of crucial importance for further miniaturization of the magnon-based spintronic devices as only short-wavelength exchange magnons allow signal processing on the nano-scale distance. Furthermore, the combination of the spin pumping and ISHE effects is the effective instrument for magnon detection beyond the wavenumber limitation of all existing methods including Brillouin light scattering spectroscopy.

Financial support by the DFG within the SFB/TRR 49, a Grant-in-Aid MEXT/JSPS, a Grant for Industrial Technology Research from NEDO, and Fundamental Research Grant from TRF, Japan is gratefully acknowledged.

\end{document}